# EXTENSION OF INTERESTING QUESTION I RAISED ABOUT 'MOMENTUM' $K_{\alpha\beta}$ OF A GEODESTIC $g_{\alpha\beta}$ AT THE J.A. WHEELER ASTRO PHYSICS SCHOOL WITH POSSIBLE NUMERICAL MODELING OF DEFORMATIONS OF LIGHT CONES

A. W. Beckwith

## ABSTRACT

I present the background of a question I asked Richard Matzner at the John Wheeler astrophysics school in Erice, at the Majorana institute of scientific culture during May 31$^{st}$ -June 8$^{th.}$ with an extension of the question to the issue of possible light cone deformation. This possibly has relevance to the stability of numerical algorithms used in modeling gravitational waves from Schwartzshield metric solutions of black holes and similar objects. This involves the so called momentum $K_{\alpha\beta}$ of a geodestic $g_{\alpha\beta}$, with its re casting in a Kerr shield co ordinates of the Schwartzshield metric. for two black holes which have anti parallel spins but which exhibit spin coupling between them.

Correspondence: A. W. Beckwith:    projectbeckwith2@yahoo.com

## INTRODUCTION

I wish first of all to thank Richard Matzner of the UT Austin relativity center for a very clear and illuminating introduction to the topic of numerical simulations of gravitational wave generation in general relativity. The topic being presented is an abbreviated introduction to part of his lecture, as a setting to an issue of 'symmetry' of the 'momentum' $K_{\alpha\beta}$ of the geodestic 'data' $g_{\alpha\beta}$ of a schwartzschild metric used in numerical simulations of gravity waves. The first part of the question which I present here was given an extensive answer by Dr. Matzner which will show up in the proceedings of the 1$^{st}$ inaugural John Wheeler School of astrophysics. The second part of the question which I am introducing here, with its extensions to light cones is something which I did not raise in the question and answer period, but which I think is extremely significant as to the issue of the behavior of light cones and their interactions in gravitational wave simulations. It was not raised, but was shown Richard Matzner, and was omitted due to wishing to keep a reasonably orderly question and answer period, and not over taxing the patience of note takers who were supposed to transcribe each and every question and the subsequent answers of the conference proceedings.

I will leave the full details of the answer to question I managed to ask Richard Matzner in the question and answer period to the fore coming conference proceedings and will, instead concentrate upon the second question which I do believe has interesting physics and should be at some time professionally investigated. This will necessitate an abbreviated introduction to the physics of this topic partly along the lines Richard Matzner made available to John Wheeler astrophysics conference , and then the brief introduction to the topic I wish to ask about , namely how light cones are affected by considerations of the 'dynamics' of the 'momentum' $K_{\alpha\beta}$ of the geodestic data $g_{\alpha\beta}$ of a schwartzschild metric. in the **Kerr-Schild** form.

## OUTLINE OF HOW TO SET UP 'MOMENTUM' OF $K_{\alpha\beta}$ OF THE GEODESTIC DATA $g_{\alpha\beta}$ OF A SCHWARTZSCHILD METRIC. ANALOGOUS TO THE KERR-SCHILD FORM.

The following is largely due to Richard Matzner[1], and will be included as supporting structure of the physics of the questions I raised which were at least considered in Erice, Sicily, in the John Wheeler astro physics school. It was part of his lecture on gravitational waves[1].

To begin with look at a **Schwarzschild solution** $g_{\alpha\beta}$ when written in a manner analogous to the **Kerr-Schild** form[2,3]. This means that one is considering a co ordinate set when considering the time derivative of $g_{\alpha\beta}$ with a non zero 'momentum' along the lines of

$$\dot{g}_{\alpha\beta} \equiv 0, \quad but \quad K_{\alpha\beta} \neq 0 \tag{1}$$

As Matzner presented it[1], this was for a modified '$dS^2$' length which he stated was for

$$dS^2 = -\left(1 - \frac{2M}{r}\right) \cdot dt^2 + 2 \cdot \left(\frac{2M}{r}\right) \cdot dt \cdot dr + \left(1 + \frac{2M}{r}\right) \cdot dr^2 + r^2 \cdot \left(d\theta^2 + \sin^2\theta \cdot d\vartheta^2\right) \tag{2}$$

This lead to, in the John Wheeler presentation Matzner made of the following abbreviations, namely

$$\beta_r \equiv \frac{2M}{r} \tag{3a}$$

and

$$\alpha^2 \equiv \frac{1}{1 + \frac{2M}{r}} \tag{3b}$$

as well as

$$t_{ks} \equiv t + 2 \cdot M \cdot \ln\left(\frac{r}{2 \cdot M} - 1\right) \tag{3c}$$

These are used in the time derivative of (assuming $K_{\alpha\beta} \neq 0$)

$$\dot{g}_{\alpha\beta} \equiv -2 \cdot \alpha \cdot K_{\alpha\beta} + \beta^k \cdot g_{ji,k} + g_{jk} \cdot \beta^k_{ji} + g_{ik} \cdot \beta^k_{ji} \tag{4}$$

This is assuming as well that

$$\dot{K}_{\alpha\beta} \equiv \left(\left(-\nabla_\alpha \nabla_\beta \cdot \alpha \neq 0\right) + \left(\alpha^3 \cdot R_{\alpha\beta} \neq 0\right)\right) = 0 \tag{5}$$

# IMPLICATIONS OF NON ZERO 'MOMENTUM' OF ($K_{\alpha\beta} \neq 0$) IN TERMS OF SPIN AND TWO INTERACTING BLACK HOLES

What already has been raised by me in a question and answer session at the Wheeler astrophysics school[1] with Richard Matzner was the following question.

## QUESTION 1:

*If there exist two black holes with anti parallel spins, then if the coupling between the spins of the two black holes is monotonically decreasing, does this mean that $K_{\alpha\beta}$ is also monotonically decreasing?*

As stated, Richard Matzners long, answer to this question will be included in full details in the John Wheeler astrophysics proceedings to be released by the Majorana institute of Erice[1]. But it leads to the following observation as discussed with both Richard Matzner, and with other scientists at the meeting. Namely that we have the opposite phenomena here occurring from electromagnetism, namely that since gravitational masses in general relativity attract as a primary effect, we need to consider secondary effects as well, namely.

## QUESTION 2

*Does the existence of a monotonically decreasing **Kerr-Schild representation of** ($K_{\alpha\beta} \neq 0$) imply deformations in the light cone of the system where two black holes may be interacting in?*

I argue this is an extremely non trivial consideration for the following reason. In an article written by Blaut, A.; et al[4], about doubly special relativity, the authors write the following in their introduction, namely that they are considering *Doubly Special Relativity theory based on the generalization of the κ-deformation of the Poincaré algebra acting along one of the null directions. They (We) recall the quantum Hopf structure of such deformed Poincaré algebra and use it to derive the phase space commutation relations. As in the DSR based on the standard quantum κ-Poincaré algebra they (we) find that the space time is non-commutative. They (We) investigate the fate of the properties of Special Relativity in the null basis: the split of the algebra of Lorentz and momentum generators into kinematical and dynamical parts, the action of the kinematical boost M+-, and the emergence of the two dimensional Galilean symmetry*

This is partly based upon an idea raised by G. Amelino-Camelia [5,6] in the general literature about doubly special relativity, and it is likely that deformations in a perceived light cone could lead to non symmetric gravity wave generation from a numerical simulation of two black holes in the configuration outlined above. In any case, this could in its own way serve as an investigation of if or not purported subtle violations of Lorentz transformations indeed have a relativistic signature which can be investigated experimentally and numerically in numerical simulations of gravitational wave physics involving spin coupling between two black holes. Purported tests of the Doubly special relativity idea have been exceedingly difficult so far[7] and this has implications as to loop quantum gravity and string theory which could serve as a falsifiable basis of determining if these theories do as they purport to do in extreme limits of the space time continuum as we know it.

## **CONCLUSIONS**

Numerous allegations as to investigations as to the structure of space time have been proposed to test if or not doubly special relativity holds water. One of the more recent has been by J. Kowalski-Glikman[8], The question so raised above would be a more phenomenon oriented way of ascertaining if or not experimental tests of Lorentz violations so purported can be extended to black hole simulation physics. In particular, as well, we should note G. Amelino-Camelia[6] has written extensively on these concerning purported minimum length ideas,

Physically realistic black hole simulations, and investigations as to deformation of light cones as to subtle interactions as proposed above could lead to falsifiable predictions and answers as to the exotic geometries proposed by string theorists and loop quantum gravity experts do as they purport to do, and allow for additional investigations of early universe models, once we know the limits of applicability of string theory and loop quantum gravity to cosmology today. This would round out ideas already in the literature presented by numerous authors who have suggested that space time metrics be investigated directly as a proving ground for falsifiable conclusions of the Doubly Special relativity idea.

## ACKNOWLEDGEMENTS

I wish to thank the Ettoire Majorana School which hoisted the 1st J.A.Wheeler school of astrophysics, and Dr. Richard Matzner for extremely lucid discussions of both gravitational wave physics, as well as standard General relativity lore pertinent to this issue, and so much more.